\documentclass[aps,prl,twocolumn,superscriptaddress,longbibliography]{revtex4-2}
\usepackage{gensymb}
\usepackage{textcomp}
\usepackage{graphicx}% Include figure files
\usepackage{subfigure}
\usepackage[percent]{overpic}
\usepackage{dcolumn}% Align table columns on decimal point
\usepackage{siunitx}
\usepackage{bm}% bold math
\usepackage{hyperref}% add hypertext capabilities
\usepackage{natbib}
\usepackage{amsmath}

\usepackage{xcolor}
\usepackage{ulem}
\DeclareSIUnit\angstrom{\text {Å}}

\newcommand{\beq}{\begin{equation}}
\newcommand{\eeq}{\end{equation}}
\newcommand{\bk}{\bm{k}}
\newcommand{\bq}{\bm{q}}

\begin{document}

% \preprint{APS/123-QED}

% \newcommand{\thistitle}{\textbf{Dynamic correlation in twisted $\text{MoTe}_2$ multi-bands: a non-perturbative renormalization group approach}}
\newcommand{\thistitle}{
Stabilizing Fractional Chern States in Twisted $\text{MoTe}_2$: Multi-band Correlations via Non-perturbative Renormalization Group
}

\title{\thistitle}

\author{Run Hou}
\affiliation{Department of Physics and Astronomy, Rice University, Houston, Texas 77005, USA}

\author{Andriy H. Nevidomskyy}
\affiliation{Department of Physics and Astronomy, Rice University, Houston, Texas 77005, USA}
\affiliation{Rice Center for Quantum Materials and Advanced Materials Institute, Rice University, Houston, Texas 77005, USA}
\affiliation{Division of Condensed Matter Physics and Materials Science,
Brookhaven National Laboratory, Upton, NY 11973-5000, USA}

 \date{\today}% It is always \today, today,
%              %  but any date may be explicitly specified

\begin{abstract}
The observation of fraction quantum Hall states in twisted MoTe$_2$ has sparked a lof of interest in this phenomenon. Most theoretical works to date rely on the brute-force exact diagonalization which is limited to the one partially occupied band. In this work, we present strong evidence that the effect of higher lying bands cannot be ignored due to strong interband interactions.
To tackle these effects, we introduce a non-perturbative driven similarity renormalization group (DSRG) method, originally developed for problems in quantum chemistry.
We apply this methodology to twisted MoTe$_2$  at fractional hole fillings of $\nu = 1/3$ and $2/3$ across a spectrum of twist angles. Our results show that at $\nu = 1/3$, the many-body  excitation energy gaps are substantially reduced compared to the one-band treatment.
For $\nu = 2/3$, we find that the
dynamic correlations stemming from interband interactions  stabilize fractional Chern insulating phases at larger twist angles, consistent with the experimental findings. 
By examining the correlated orbitals and their single-particle topological features, we demonstrate that this stabilization at higher twist angles arises predominantly from the dynamic correlations, rather than conditions on the single-particle  quantum geometric tensor. %of ideal quantum geometric conditions.
\end{abstract}
% covered at length in the main body of the article. 
% \begin{description}
% Secondary publications and information retrieval purposes.
% You may use the \texttt{description} environment to structure your abstract;
% use the optional argument of the \verb+\item+ command to give the category of each item. 
% \end{description}
% \end{abstract}

\maketitle

% \tableofcontents

\textit{Introduction}. The recent experimental discoveries of fractional Chern insulator (FCI) states in twisted bilayers of transition metal dichalcogenides (TMDs) \cite{FCICai2023,FCIPark2023,FCIZeng2023,FCIJi2024,FCIWang2025} mark a significant advancement in identifying exotic strongly correlated states in real materials. Subsequent observations of superconductivity \cite{SCGuo2025,SCXia2025} have further heightened interest, providing tunable platforms to explore the intricate interplay between topology and strong electronic correlations.

Similar to Landau levels, the topological flat bands in twisted MoTe$_2$ (tMoTe$_2$) host integer and fractional quantum Hall (FQH) phenomena, notably without the necessity of applying external magnetic fields. Due to intrinsic spin-orbit coupling, spins and valleys are locked, ensuring a $U$(1) charge conservation for each spin-valley species. Strong exchange interaction results in quantum Hall ferromagnetism, fully polarizing systems at small twist angles, as confirmed by theoretical investigations \cite{VPED,LiangFuED,JiabinMulti,VPtheoryQiu}. Within this fully spin-valley polarized sector, numerical studies have identified FCI and generalized Wigner crystal (gWC) phases at fillings of $\nu = 1/3$ and $2/3$ \cite{VPED,LiangFuED,JiabinMulti,XuPNAS,TMDmodel,PressureED,displacementED}, with the gWC characterized by charge density wave (CDW) ordering at the mini-moiré Brillouin zone's $K$ point.

Theoretical investigation of twisted TMDs is complicated by the long-range nature of the Coulomb interaction and the fact that modeling the FCI states requires techniques beyond the mean-field Hartree--Fock treatment that has been used routinely to study correlated phases in other moir\'e systems.
Thus, most theoretical approaches employed so far have  utilized exact diagonalization (ED), which inherently suffers from exponential computational complexity, limiting scalability. Typically, these models are projected onto the lowest partially occupied moir\'e band prior to performing the ED. However, the validity of such a projection is questionable, given that interaction strengths often exceed the band gap and one thus expects interband dynamics to be important. Several multi-band analyses have attempted to address these limitations \cite{JiabinMulti,XuPNAS,mixingLiangFu}, but have been limited to relatively small system sizes by the computational complexity. 

In this paper, we adopt the strategy of a non-perturbative %renormalization group (RG) 
method called similarity renormalization group (SRG), originally proposed by Wilson, G\l{}azek and Wegner \cite{Wilson1993,Wegner1994} as a way of downfolding the multi-band models into effective single-band Hamiltonians. Pragmatically speaking, this technique allows one  to overcome the exponential wall encountered in ED of multi-band systems by employing a method with lower polynomial complexity. Crucially, the resulting single-band Hamiltonian encapsulates non-pertubatively the \textit{dynamic correlations} arising from the interband interaction, thus serving as a significant improvement upon the simple projection onto the lowest moir\'e band.
%such that we overcome the exponential wall by partially treating the system with a method that has lower polynomial complexity.

Traditional SRG methodologies rely on normal ordering and Wick's theorem for single Slater determinant reference states, which is inadequate for describing strongly correlated FCI states exhibiting multi-Slater determinant characteristics. Analogous challenges are recognized in strongly correlated Hubbard and Anderson lattice models, often manifesting as divergent RG flows. Additionally, SRG formulations typically involve solving stiff systems of ordinary differential equations, presenting numerical difficulties. These issues have been recently addressed within quantum chemistry community through the development of the multi-reference driven similarity renormalization group (MR-DSRG) method \cite{DSRG2014,DSRG2,DSRGexcited,annurev:DSRG}, which combines multi-Slater determinant reference states with the generalized normal ordering (GNO) expansions \cite{GNOMukherjee1995,GNO1997,GNOMUKHERJEE1997561,GNO2013}, and converts differential equations into integral equations for enhanced stability.
Throughout this work, we adopt multi-Slater determinants as the reference state, and will refer to our method simply as driven SRG (DSRG) for the sake of brevity. Extensive studies demonstrate DSRG's robustness and accuracy in capturing molecular dissociation potential energy curves and excitation energies \cite{DSRGexcited,DSRGex2}. To date, this method has not been applied to extended periodic systems, to the best of our knowledge, and the present work showcases the power of the DSRG method.

Applying DSRG to tMoTe$_2$ systems, we show the importance of band mixing effects in moiré materials, demonstrating that \textit{dynamic} interband correlations significantly affect the stability of the FCI states. Crucially, we demonstrate that band projection alone (onto the lowest moir\'e band)  is insufficient to fully characterize the phase diagrams and excitation properties. Using DSRG, we investigate the stability of the FCI and CDW states within the phase diagram as a function of twist angle and top gate distance, and correlate these phases with the topology and quantum metric of the model.
Our DSRG framework additionally provides a natural bridge connecting topological multi-band systems and generalized Anderson lattice models, highlighting its versatility and broad applicability of DSRG to correlated electron systems.

\textit{Model}. The one-body hole Hamiltonian of twisted TMDs (in a given spin-valley) is expressed by a continuum model,

\begin{equation}
    H_{\uparrow} = \begin{pmatrix}  \frac{\hbar^2 (-i \nabla -\kappa_+)^2}{2m} + V_+(\bm{r}) & T(\bm{r})\\
    T^\dagger(\bm{r})&\frac{\hbar^2 (-i \nabla -\kappa_-)^2}{2m} + V_-(\bm{r})\end{pmatrix}, 
    \label{eq:model}
\end{equation}
where $\bm{\kappa}_\pm = \frac{4\pi}{3 a_M} (-\sqrt{3}/{2}, \mp {1}/{2})$ are vectors of the two $K$ point in moir\'e Brillouin zone, with moir\'e lattice constant $a_M = a_0/[2\sin{(\theta/2)}]$. $V_{\pm}(\bm{r})$ are the moir\'e potentials, and $T(\bm{r})$ is interlayer tunneling. Both potentials are expanded to lowest harmonics: 
%$V_{\pm}(\bm{r}) = \sum_{i=1,3,5}2V\cos[\bm{G}_i\cdot \bm{r}\pm \phi]$, and $T(\bm{r}) = w\left(1+e^{-i \bm{G}_2 \cdot \bm{r}} +e^{-i \bm{G}_3\cdot \bm{r}}\right)$, 
\begin{align}
V_{\pm}(\bm{r}) &= \sum_{j=1,3,5}2V\cos[\bm{G}_j\cdot \bm{r}\pm \phi], \nonumber \\  
T(\bm{r}) &= w\left(1+e^{-i \bm{G}_2 \cdot \bm{r}} +e^{-i \bm{G}_3\cdot \bm{r}}\right)
\end{align}
where  $\bm{G}_j = \frac{4\pi}{\sqrt{3}} \left(\cos [\pi(j-1)/3] , \sin [\pi(j-1)/3]\right)$ are the reciprocal lattice vectors.  The actual values of the model parameters are extracted from DFT calculations \cite{TMDmodel}, we choose them to be $(a_0,V,w,\phi,m) = (0.352~\si{\angstrom},~ 20.8~\text{meV},~ -23.8~\text{meV},~-107.7\degree,~ 0.60~m_e)$, such that the FCI states can be obtained around the experimentally observed twist angles in MoTe$_2$~\cite{FCICai2023,FCIPark2023,FCIZeng2023,FCIJi2024,FCIWang2025}. The opposite spin/valley one-body Hamiltonian can be obtained by applying time-reversal operation. Throughout the paper, we will assume the ground state to be fully spin polarized, thus focusing on one spin-valley in Eq.~\eqref{eq:model}. 

We project the interactions into the lowest \textit{three} bands. The interacting Hamiltonian is
\begin{equation}\label{eq:interaction}
    H_I = \sum_{\substack{\bm{q}\\ \sigma_1, \sigma_2 } } \frac{1}{2N\Omega_c}V(\bm{q}) :\tilde{\rho}_{\sigma_1}(\bm{q}) \tilde{\rho}_{\sigma_2}(-\bm{q}):,
\end{equation}
where $\tilde{\rho}_{\sigma}(\bq ) = \sum_{\bk}\sum_{mn} M_{mn;\sigma}(\bk,\bq) c_{m\sigma}^\dagger(\bk+\bq)c_{n\sigma}(\bk)$ is momentum-space projected density operator dressed by the form factor $M_{mn;\sigma}(\bk,\bq)$ (see \textit{End Matter}), and the interacting Hamiltonian is normal ordered with respect to the physical vacuum. %\anc{Does the bar over $\bar{\rho}$ imply that it's somehow renormalized? Can I write $\bar{\rho}$ in a standard way via non-interacting operators, $\bar{\rho}(q)=\sum_k c^\dagger_{k+q}c_k$?}
The double-gated screened Coulomb interaction is taken, as usual, to be
\begin{equation}
    V(\bm{q}) = \frac{2\pi e^2}{\epsilon}\frac{\tanh (|\bm{q}|d_g/2)}{|\bm{q}|},
\end{equation}
where $d_g$ is the gate distance, and the dielectric constant is $\epsilon = 10$ through out the work~\footnote{Here, we do not consider any double counting subtraction scheme, as the actual effective potential is unknown.}.

\textit{Methods: SRG and DSRG.} The essence of the SRG method consists in finding a desired similarity transformation of the Hamiltonian. Considering an arbitrary many-body unitary rotation $e^{A}$ on a many-body reference state $\vert \Phi  \rangle$, the transformation modifies the state to $\vert \Psi\rangle = e^{A}\vert\Phi\rangle$, where $A$ is an anti-hermitian operator. Or, switching to the Heisenberg picture, the transformed Hamiltonian becomes 
\beq
\overline{H} = e^{-A} H e^{A}.
\eeq 
Diagonalizing $\overline{H}$ would give references $\Phi$, and the true wave function in the original basis can be viewed as $\Psi$. Here, the anti-hermitian operator $A$ is iteratively found such that the off-diagonal terms which couple the lowest band (so-called \textit{active space}) to the other bands are eliminated $\overline{H}_{\text{od}} \rightarrow 0$. We use Baker–Campbell–Hausdorff (BCH) expansion to evaluate the unitary transformation of the operators. However, the expansion generally does not terminate as we choose unitary operators to rotate the many-body Hamiltonian. To obtain a closed system of equations, the higher rank terms beyond three-body terms are truncated, and the three-body terms themselves are approximated using one- and two-body density matrices following GNO rules (the fully normal-ordered three-body terms are neglected)~\cite{GNOMukherjee1995,GNO1997,GNOMUKHERJEE1997561,GNO2013}.  This scheme is called DSRG(2) \cite{DSRG2},  resulting in an effective Hamiltonian with up to two-body terms, which can be shown, via the BCH expansion, to be of the  following form:
\begin{equation}\label{eq:bch}
    \overline{H} = \sum_{k =0} B_k, \text{ where } B_k = \frac{1}{k} [ B_{k-1} , A ]_{0,1,2}
    %\label{eq:recursion}
\end{equation}
can be evaluated recursively as per the above starting with  $B_0 = H$, and the subscripts 0,1,2 represent keeping up to normal ordered two-body terms within the GNO framework. The commutator and normal ordering have been calculated using the automatic derivation package \textsc{Wick\&d} \cite{wickd}. The recursive expansion is terminated when the norm of the operator difference is smaller than $10^{-12}$.  
The DSRG uses a self-consistent procedure to determine the operator $A$, and details can be found in the supplemental materials. The calculation bottle neck is from the 
Eq.~\eqref{eq:bch},
%Eq.~\ref{eq:recursion} , 
where the complexity scales as $N_{com}\times\mathcal{O} (N^4_{tot} N^2_{act})$, where $N_{com}$, $N_{tot}$, $N_{act}$ are the numbers of commutators, total orbitals, and active space orbitals, respectively. 

The calculation procedure is briefly summarized here. Initially, we exactly diagonalize the Hamiltonian projected to the lowest band, and compute the ground-state reduced one-body and two-body density matrices $\gamma_1$, $\gamma_2$ which will be then used in the GNO expansions. We then evaluate the DSRG(2) effective Hamiltonian using Eq.~\ref{eq:bch} and GNO. Finally, the effective Hamiltonian  $\overline{H}$ is exactly diagonalized in the active space consisting of the lowest renormalized band to obtain the ground and excited states. In order to get the observables, we evaluate the operator expectations using the expansion similar to that in Eq.~\eqref{eq:bch}, since the true expectation value of some operator $\widehat{O}$ 
%$\langle \Psi\vert \widehat{O}\vert\Psi\rangle$ 
has to be evaluated in the many-body state $e^{A}\vert\Phi\rangle$, which eventually boils down to evaluating the renormalized operator $\widehat{\overline{O}}$ in the reference state $\langle \Phi \vert e^{-A} \widehat{O} e^{A} \vert\Phi\rangle = \langle \Phi \vert  \widehat{\overline{O}}\vert\Phi\rangle$.

\begin{figure}[t!]
    \centering
    \includegraphics[width=\linewidth]{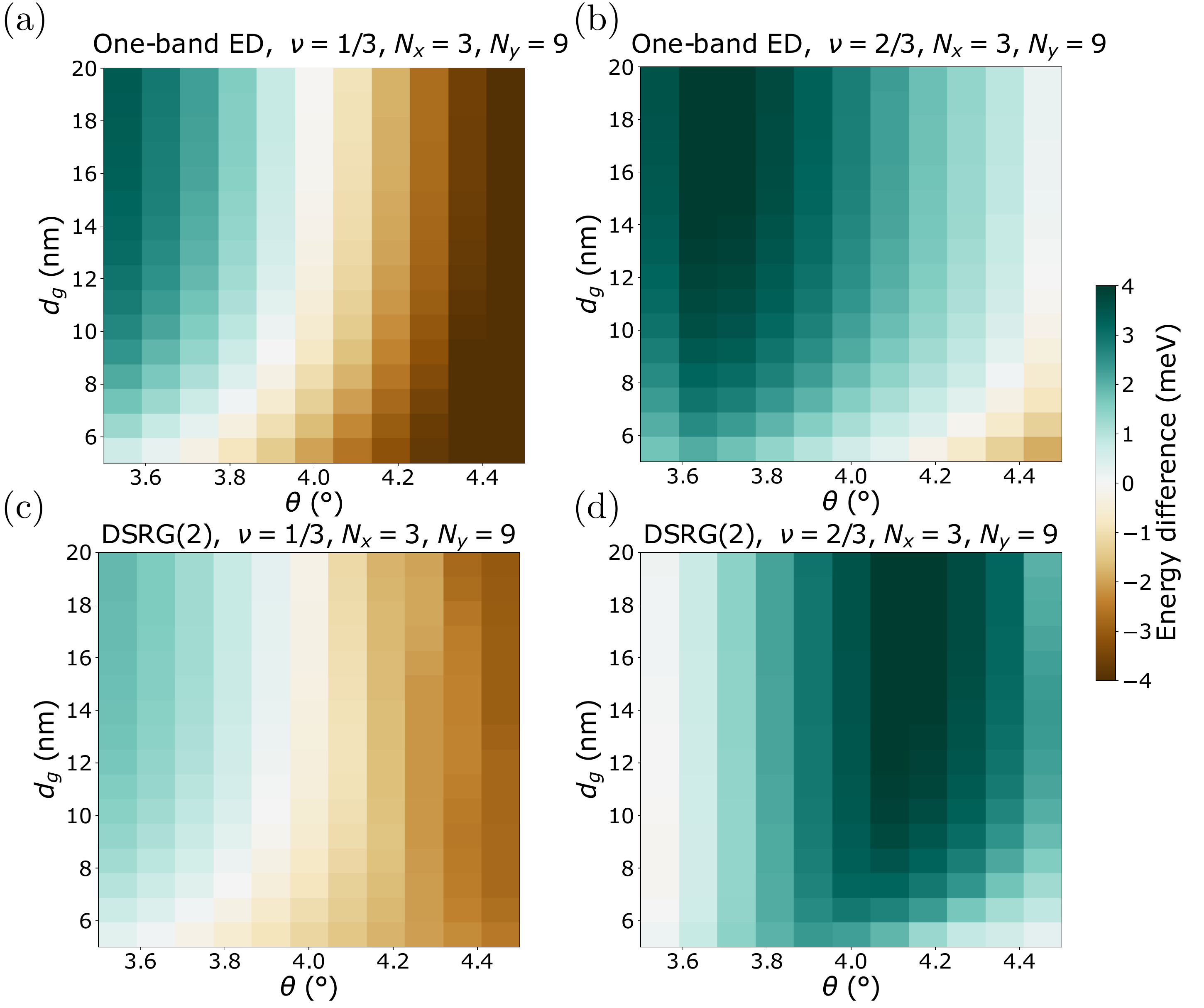}
    \caption{The minimum energy difference between CDW and FCI states, $E_{CDW} - E_{FCI}$, as a function of gate distance $d_g$ and twist angle $\theta$. Panels (a) and (b) show one-band ED results at fillings $\nu = 1/3$ and $2/3$, respectively. Panels (c) and (d) display the corresponding DSRG(2) results at these fillings.}
    \label{fig:gap}
\vspace{-2mm}
\end{figure}

\textit{Stability of $\nu=1/3$ and $2/3$ FCI in tMoTe$_2$}.
First, we benchmark the DSRG(2) method for two lowest bands on a discretised  $2\times 6$ k-mesh against the full 2-band ED (24 k-orbital), showing excellent agreement (see Fig.~\ref{fig:benchmark} in \textit{End Matter}).
Having performed the benchmarks, we now aim to solve the three-band Hamiltonians of tMoTe$_2$, where the Hilbert space is too large to perform the ED. The phase diagrams are calculated using DSRG(2) in $3\times 9 $ clusters at fillings $\nu = 1/3$ and $\nu = 2/3$ varying the gate distances in the relevant interval $d_g \in [5,20]\text{nm}$ and twist angles $\theta \in [3.5\degree,4.5\degree ]$, where the lowest and second bands have the opposite Chern number $\text{Ch} = \pm1$. Fig.~\ref{fig:gap} shows the minimum energy differences between the CDW states and FCI states $E_{CDW} -E_{FCI}$.
The negative value 
%in the band projected ED 
indicates the onset of the CDW order.
%, which we verify by evaluating the structure factor $S(\bm{q})$ (see \textit{End Matter}) shown in Fig. \ref{fig:DSRGcompare}c,d. 

At $\nu = 1/3$, comparing DSRG(2) with the band projected ED results, the overall excitation gaps of FCI states are smaller than those of the band projected ED, as shown in the Fig.~\ref{fig:gap}(a,c). This is an indication that the interband interactions destabilize the FCI states. For $\nu = 2/3$, the false color plots of the energy gap in Fig.~\ref{fig:gap}(b,d) show that the DSRG(2) excitation gaps peak at larger twist angles compared with the one-band ED. In other words, the interband dynamic correlations which are captured by DSRG(2) destroy the tendency toward the CDW order, making the FCI states more stable at $\nu = 2/3$. This is consistent with the experimental observations \cite{FCICai2023,FCIPark2023} of FCI at twist angles  larger than the optimum angle $\sim3.6\degree$ predicted by band-projected ED.
%, highlighting the advantage of our DSRG method.
%{Although the model twist angles do exactly corresponding to the angles of experimental devices, experiments observe FCIs in larger twist angles \cite{FCICai2023,FCIPark2023} compared with optimal angles $3.6\degree$ in band-projected ED calculations, suggesting the dynamics of higher bands have impacts on the optimal twist angles in realist materials.} 

\begin{figure}[t]
    \centering
    \includegraphics[width=\linewidth]{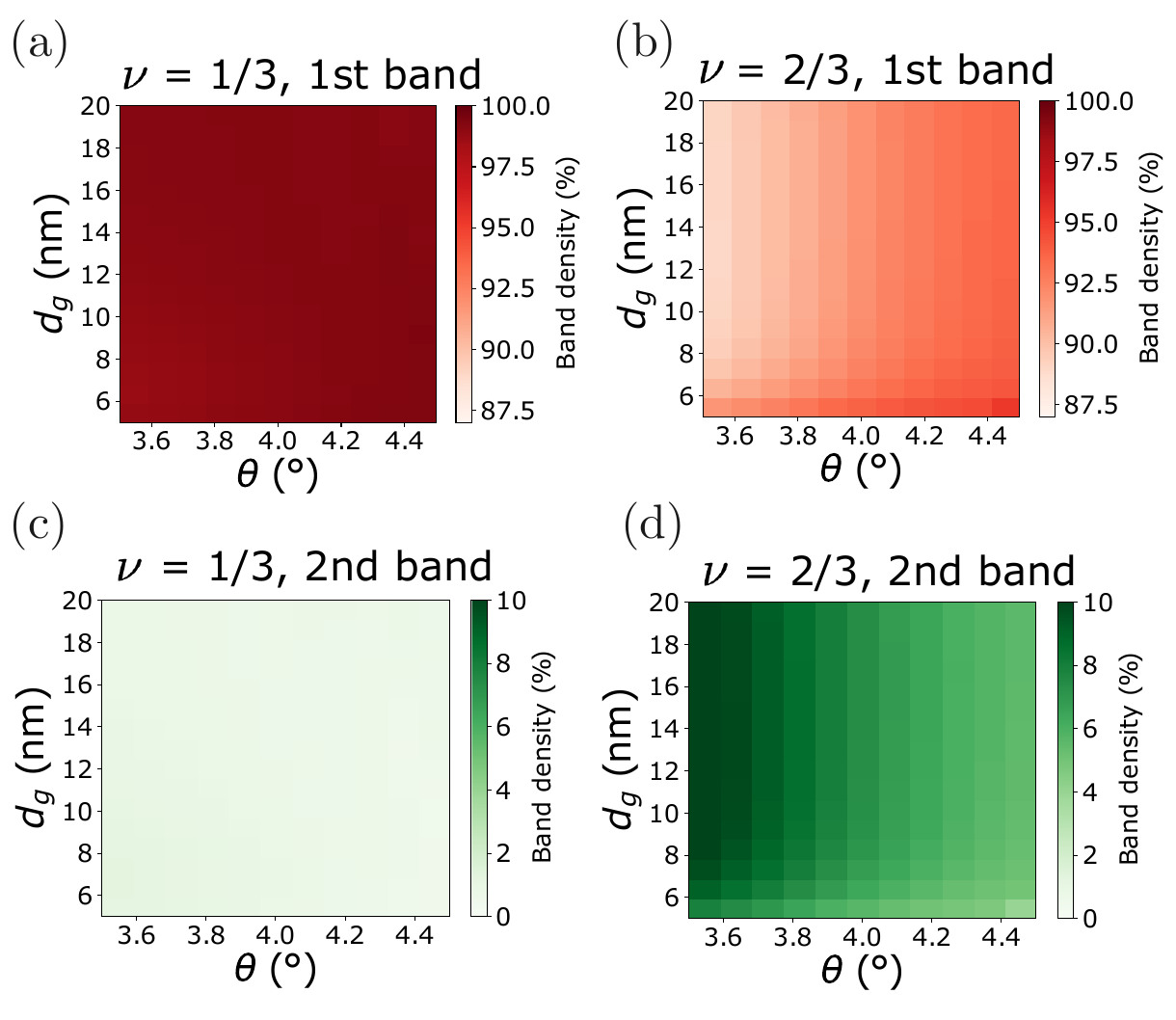}
    \caption{Band occupation analysis for ground state $e^{A} \vert \Phi'\rangle$, as a function of gate distance $d_g$ and twist angle $\theta$. Panels (a) and (b) show first-band occupation percentages at fillings $\nu = 1/3$ and $2/3$, respectively. Panels (c) and (d) display the corresponding second-band results at these fillings. }
    \label{fig:BandOccupation}
\vspace{-2mm}
\end{figure}

\textit{Band mixing}. In order to understand how much the band mixing effect is, we evaluate the $n^\text{th}$ band occupation expectation values $N_n = \sum_{\bm{k}}\langle c^\dagger_n(\bm{k})c_n(\bm{k})\rangle$ where $c_n^\dagger(\bm{k})$ is the  electron creation operator in the $n^\text{th}$ lowest band. Following the DSRG(2) procedure described above, we can evaluate the renormalized expectation values $\overline{N}_n$, for which the details can be found in \textit{End Matter} Eq.~\ref{eq:density operator}. We plot the occupation percentage of the first  and second band in Fig.~\ref{fig:BandOccupation}. At $\nu = 2/3$, the lowest band occupation can be around 89\% - 95\%, and the second band occupation is up to 9\%, while at $\nu = 1/3$, the band mixing effect is negligible. Our numerical results suggest that the band mixing effect is negligible beyond the third band, and the electrons chiefly occupy the lowest band. %the main occupation of the electrons are still within the first band. 
Our results are similar to those obtained recently using neural network wave functions~\cite{li2025deeplearningshedslight}. At first glance, this would suggest the band projected ED is a good approximation. However, while the occupation of the second band  is small, the dynamic correlation can change a large excitation energy as shown in Fig.~\ref{fig:gap}.  

\begin{figure}
    \centering
    \includegraphics[width=\linewidth]{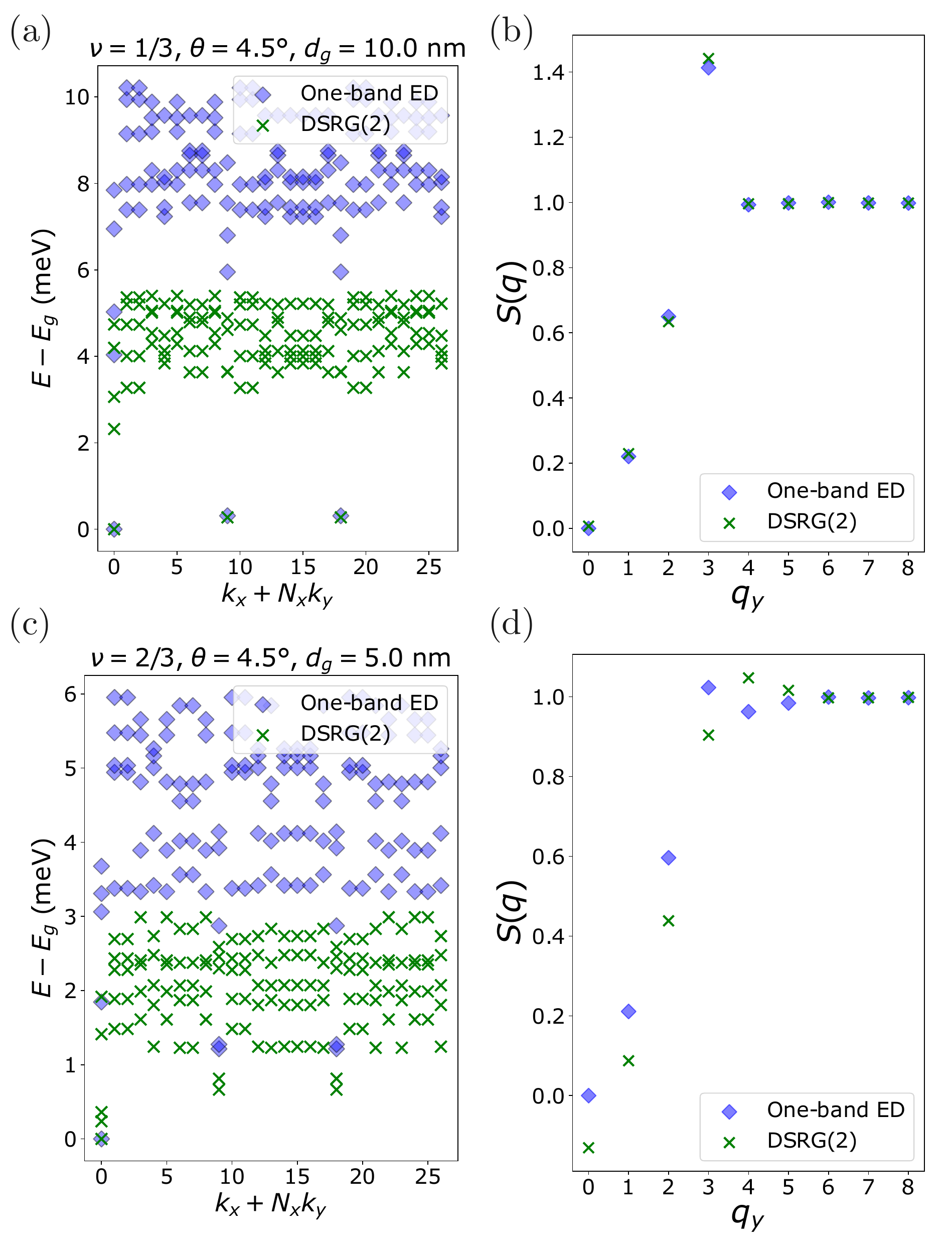}
    \caption{Excitation energies $E-E_g$ and structure factors $S(\bq)$ in $\Gamma-K$ line at fillings $\nu = 1/3$ and $\nu = 2/3$, when $\theta = 4.5\degree$ and $d_g = 10~\text{nm}$. The moir\'e $K$ point corresponds to $q_y = 3$ in the structure factor plots.}
    \label{fig:DSRGcompare}
\vspace{-2mm}
\end{figure}

\textit{Correlation effects}. 
To further verify the nature of the FCI state, we compute the particle entanglement spectrum (PES), see \textit{End Matter} and see Fig.~\ref{fig:benchmark}b. Another quantity of particular interest is the structure factor $S(\bm{q})\equiv  \frac{1}{N_e} \langle \sum_{i,j} e^{-i\bm{q}\cdot(\bm{r}_i - \bm{r}_j)}\rangle$ evaluated in the many-body ground state (see \textit{End Matter}). 
Recently, the seminal works \cite{TopologicalBound2024,zaklama2025structurefactortopologicalbound} demonstrated that in a system of Chern bands, there is a rigorous lower bound, namely $K_{\alpha \beta} \geq \text{Ch}$, where $K_{\alpha \beta}$ is proportional to the structure factor: $S(\bm{q}) = \frac{q^\alpha q^\beta}{2\pi} K_{\alpha \beta}+\ldots$.
%\anc{Run: why $\approx$ sign? Can't we define $K_{\alpha\beta}$ rigorously with an equal sign above?}
 In Fig.~\ref{fig:DSRGcompare}, we pick two representative cases at $\theta = 4.5\degree$ in the phase diagrams to compare the structure factor along the $\Gamma-K$ line before and after renormalization. 

At $\nu = 1/3$ in Fig.~\ref{fig:DSRGcompare}(a), the energy gap between the (approximately degenerate) three-fold ground state and the  excitation continuum gets lowered in SRG. The ground state is actually a CDW, as indicated by the peak at $\bm{q}=K$ point in the structure factor $S(\bm{q})$ in Fig.~\ref{fig:DSRGcompare}(b) -- similar behavior to one-band ED results, except $S(K)$ gets even higher, featuring an enhanced CDW order.
%At $\nu = 1/3$ in Fig.~\ref{fig:DSRGcompare}(a),\rh{ \sout{the ground state is a FCI}(I have a question here)}, with the energy difference to the CDW state not affected significantly by the SRC, however the energy gap between the (approximately degenerate) three-fold ground state and the  excitation continuum gets lowered in SRG. 
Even though the ground states hybridize little with other bands (see Fig.~\ref{fig:BandOccupation}a-c), the excited states are still affected by the interband dynamic correlations. 
%In Fig.~\ref{fig:DSRGcompare}(b), the structure factor also keeps the similar behavior with one-band ED results, but $S(K)$ gets even higher, featuring a enhanced CDW order.
%
For the case we pick at $\nu = 2/3 $, the one-band ED results find the lowest energy state at $\Gamma$, and have lower excitations at $K$ and $K'$ points. At first glance, one could suspect the ground state is a CDW state, since the structure factor has a small peak at the $K$ point as shown in Fig.~\ref{fig:DSRGcompare}(d). However, this peak disappears after the SRG. The renormalized excitation energy spectrum in Fig.~\ref{fig:DSRGcompare}(c) shows three-fold nearly degenerate ground states, suggesting the stabilization of FCI due to interband dynamic correlation. Intriguingly, $S(\bm{q})$ has negative values at $\Gamma$ point, which could indicate the formation of the attractive interaction.

\textit{Single-particle topology and quantum geometry}.
In addition to the band Berry curvature which contains the topological information, it is well known that a more general quantity is the Fubini-Study metric  $\mathcal{B}_{ij} = \langle \partial_{k_i} u_{\bk} |(1-|u_{\bk} \rangle \langle u_{\bk}|)| \partial_{k_j}u_{\bk}\rangle$ defined for the quantum state $|u_{\bk} \rangle$~\cite{Resta2011,Yu2025}. The real and imaginary part of this metric, $\mathcal{B}_{ij}=g_{ij} + i f_{ij}/2$ are the quantum metric $g_{ij}(\bk)$ and the Berry curvature $F(\bk)=f_{12}$. 
It is well established that these two quantities satisfy the inequality $\mathrm{Tr}(g) \geq |F|$, which is saturated in the flat Landau levels in 2-dimensional electron gas~\cite{tracecondition}. This is known as the \text{ideal} condition on the quantum metric, which in particular was shown to also hold in the chiral limit of the twisted bilayer graphene\cite{ctbg2020,ctbg2021,ctbg2022}. We now investigate whether this condition also holds in the regime where FCI is stable in twisted MoTe$_2$ (see \textit{End Matter} for  details).

\begin{figure}[t]
    \includegraphics[width=\linewidth]{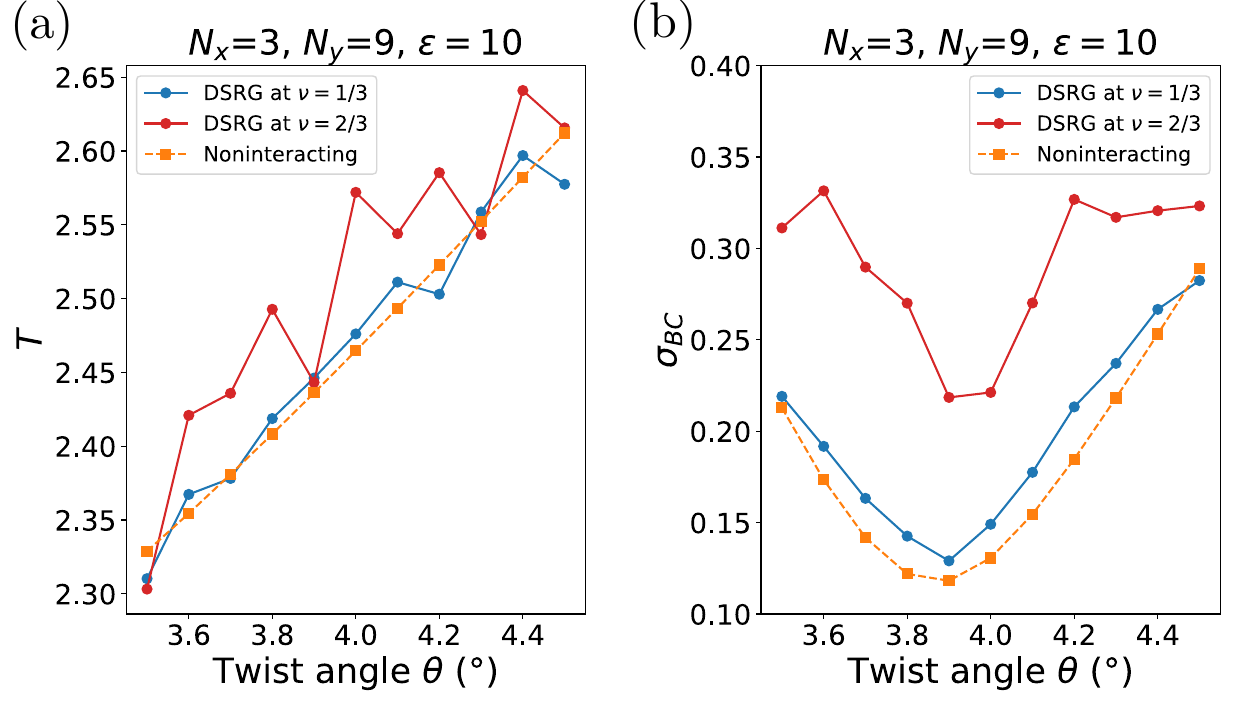}
    \caption{(a) the trace condition violation $T$ and (b) the Berry curvature standard deviation $\sigma_{BC}$  as a function of twist angle $\theta$ with fixed interlayer spacing $d_g = 10~\mathrm{nm}$.}
    \label{fig:berry}
\vspace{-2mm}    
\end{figure}

The violation of the trace condition $T = \int_{\sf{BZ}} \mathrm{d}^2 k\, \left(\mathrm{Tr}(\overline{g}_{ij}(\bk)) - \overline{F}(\bk) \right)/2\pi$ is plotted in Fig.~\ref{fig:berry}(a), from which we observe that the $\nu=1/3$ data points fall  close to the non-interacting curve, while the $\nu=2/3$ results are more ragged but follow the same general pattern, with $T$ increasing with the twist angle. Notably, no clear feature is seen at $\theta\approx 4.1^\circ$ where the FCI state is most stable (c.f. Fig.~\ref{fig:gap}d)

Another way to quantify the proximity to the chiral limit is by measuring the deviation of the Berry curvature from the constant value. To this end, we plot the standard deviation (STD) $\sigma_{BC}$ in Fig.~\ref{fig:berry}(b). The overall trend is similar in $\sigma_{BC}$ and $T$, where 
both quantities are larger at $\nu = 2/3$, while the lower filling $\nu=1/3$ does not deviate too much with respect to the original projected band model. That is consistent with the fact that band mixing is larger at the higher hole filling $\nu = 2/3$, as evidenced by Fig.~\ref{fig:BandOccupation}.
Therefore, we conclude that at $\nu = 2/3$ the improved stability of the FCI state at larger twist angles $\theta \approx 4.1^\circ$ is not due to the improvement of the ideal chiral conditions on the quantum metric, but rather is the result of the  dynamic inter-band correlations.

%The DSRG method effectively serves as a post-ED calculation to address missing dynamic correlations within projected active spaces, maintaining computational tractability. Extending this method by integrating alternative active space solvers, such as density matrix renormalization group (DMRG), could further enhance the accessible system size. Additionally, implementing orbital optimization would refine the basis representation of the full Hamiltonian, potentially improving DSRG results.

\textit{Discussion}. We have introduced the DSRG method, first developed in the quantum chemistry community, and have successfully applied it to the strongly correlated problem of FCI phases in twisted MoTe$_2$. DSRG addresses missing dynamic correlations within the projected active space, while maintaining computational tractability. 
Notably, DSRG(2) calculations can be fully executed in momentum space, preserving essential topological information. Moreover, DSRG remains applicable even when Wannier orbitals cannot be maximally localized due to band topology.

Recent studies~\cite{TopoAnderson,xie2025kondolatticephenomenologytwistedbilayer} have introduced topological Anderson models in twisted TMD systems. While dynamic mean-field theory (DMFT) approaches typically simplify or neglect long-range interactions -- potentially obscuring topological characteristics -- the DSRG method inherently accommodates all interaction terms, thus retaining full topological fidelity within larger, yet numerically manageable, system sizes compared to traditional ED methods. Extending this method by integrating alternative active space solvers, such as density matrix renormalization group (DMRG), could further enhance the accessible system size and is left to future studies.

Another promising application of DSRG(2) in solid-state systems involves extracting effective attractive interactions at various fillings in twisted TMDs. Such insights are valuable for elucidating strongly correlated superconductivity phenomena associated with flat bands, as well as for investigating other exotic fractionalized phases.

\textit{Acknowledgment.} We thank Fang Xie for helpful discussions. 
%This work was supported by the Department of Energy under the Basic Energy Sciences award no. DE-SC0025047. 
This work was supported by the U.S. Department of Energy (DOE) Office of Basic
Energy Sciences (BES), Division of Materials Sciences and Engineering under contract No. DE-SC0012704 with Brookhaven National Laboratory.
The computing resources at Rice University were supported in part by the Big-Data Private-Cloud Research Cyberinfrastructure MRI-award funded by NSF under grant CNS-1338099 and by Rice University's Center for Research Computing (CRC).

\section*{End Matter}

\textit{Interactions in tMoTe$_2$}. The double-gate screened Coulomb interaction is given by
\begin{equation}
    V(\bm{q}) = \pi d_g^2 V_d \frac{\tanh (d_g|\bm{q}|/2)}{d_g|\bm{q}|/2}, \quad V_d = \frac{e^2}{4 \pi \epsilon \epsilon_0 d_g}.
\end{equation}
In our calculations, we set the distance $d_g$ between the gates to be within the range of $5~\text{nm} \leq d_g \leq 20~\text{nm}$, and the dielectric constant $\epsilon^{-1}$ to be 0.1.
The projected momentum-space density operator in the low-energy bands are constructed by form factors $M_{mn;\sigma}(\bm{k},\bm{q})$,
    \begin{equation}
    \begin{split}
        \tilde{\rho}_{\sigma}(\bm{q}+\bm{G}) &= \sum_{\bm{k}}\sum_{mn}  M_{mn;\sigma}(\bm{k},\bm{q}+\bm{G})\\
        &c^\dagger_{m\sigma}(\bm{k}+\bm{q}+\bm{G})c_{n\sigma}(\bm{k}) \\
        M_{mn;\sigma}(\bm{k},\bm{q}) &=\sum_{\bm{Q}_l}  u^\ast_{\bm{Q}_l m;\sigma}(\bm{k}+\bm{q})u_{\bm{Q}_l n;\sigma}(\bm{k}),
    \end{split}
\end{equation}
where $m$, $n$ are band indices, and $\sigma$ is the valley/spin indices.
$u_{\bm{Q}_l n;\sigma}(\bm{k})$ represents the periodic part of Bloch functions of the low-energy bands.

\begin{figure}[t!]
    \centering
    \includegraphics[width=\linewidth]{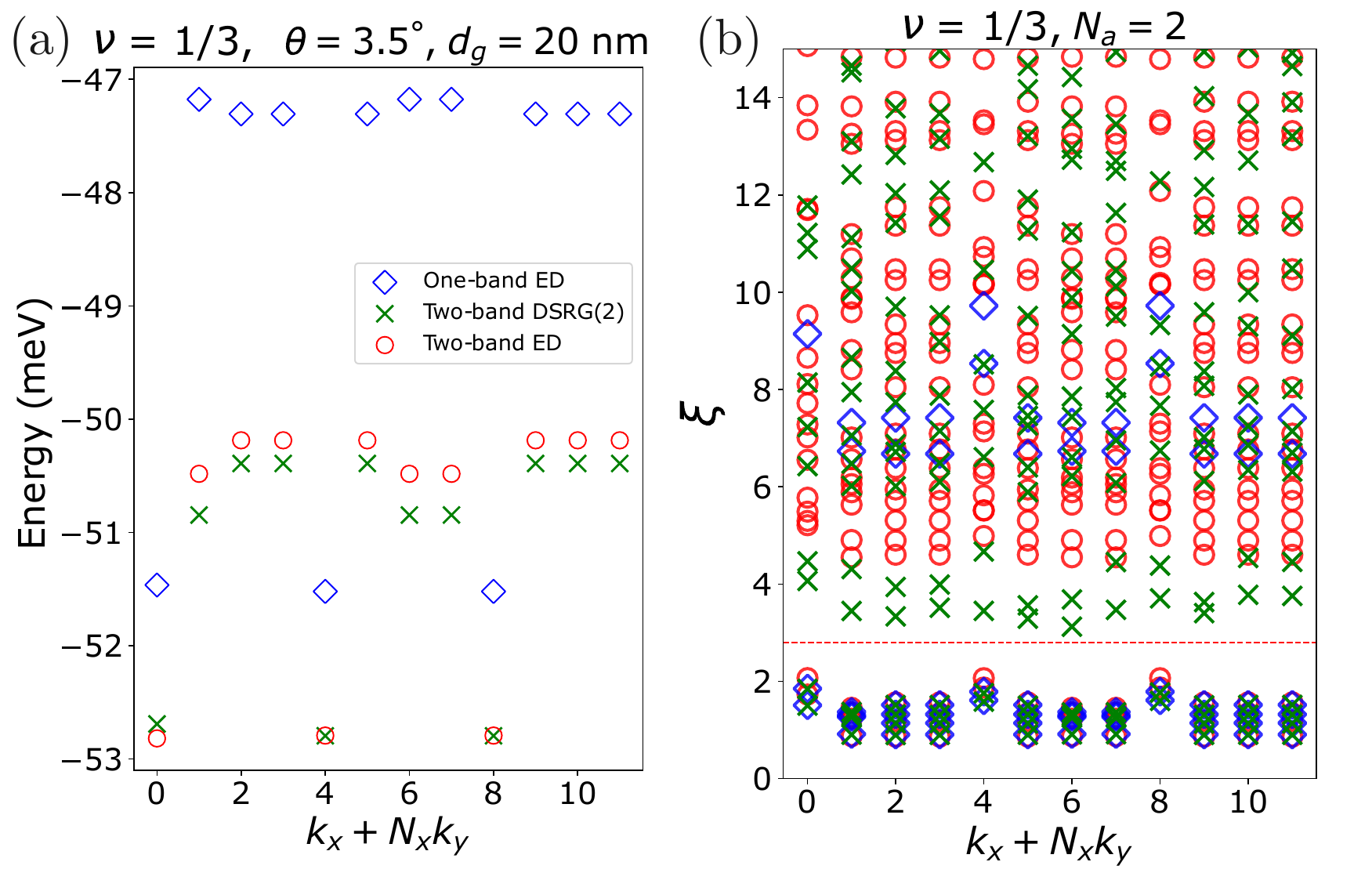}
    \caption{Numerical results of two-band DSRG(2) vs. one-band and two-band ED at twist angle $\theta = 3.5\degree$, filling fraction $n=1/3$, and gate distance $d_g = 20 ~\text{nm}$ in a $2\times 6$ cluster. (a) The first energy eigenvalues of each momentum. The DSRG(2) method captures the energies from interband interactions. (b) The particle-entanglement spectrum $\xi$ of three calculations. The counting of states below the dashed line is 42, which is consistent with the expectation from the generalized Pauli principle for the FCI state~\cite{Regnault2011}.}
    \label{fig:benchmark}

\vspace{-2mm}
\end{figure}

\textit{Benchmarking against ED}. To begin with, we benchmark the DSRG(2) method with ED in a small $2\times6$ cluster using a two-band $\text{tMoTe}_2$ Hamiltonian. In Fig.~\ref{fig:benchmark}(a), the comparison of the energy eigenvalues from different methods shows that DSRG(2) is able to capture not only the ground state energies but also the excitation energies, while the one-band projected ED fails to obtain these features, showing the significance of interband correlations. Notably, our DSRG(2) only needs the Hilbert space dimension of the same size ($\sim\!40$) as the one-band ED, whereas the full two-band ED would need $\sim\!900$ states, showing the computational advantage of our algorithm.

\textit{Characterizing the FCI state.}
A useful tool often used to characterize FCI states is the particle entanglement spectrum (PES), computed based on partitioning the system into two parts based on particle number $N = N_A+N_B$ (rather than spatial regions).
The wavefunction at a given momentum $\bm{k}$ can be written in the form $\Psi(\bm{k}) = \sum C_{ij} |\phi_i, N_A\rangle \otimes |\phi_j, N_B\rangle$, where the sum is over the direct-product states of the two subsystems. 
The coefficient matrix $C_{ij}$ carries the information about entanglement, which can be extracted via the singular-value decomposition $C_{ij} = \sum_\alpha\exp(-\xi_\alpha/2)  U_{\alpha i} V_{\alpha j}$, with $\xi_\alpha$ being the PES eigenvalues. 
These are plotted in Fig.~\ref{fig:benchmark}(b) for $N_A=2$ at filling $\nu=1/3$, from which it is evident that the leading PES eigenvalues below the dashed line are separated by a gap from the rest. 
In a $\nu=1/3$ FCI state, the number of these low-lying $\xi_\alpha$ is expected to follow the generalized Pauli principle~\cite{Regnault2011}: $N_{\text{PES}}=N_xN_y(N_xN_y-2N-1)!/[N!(N_xN_y-3N)!]$, which is equal to 42 in the $2\times 6$ cluster. That is precisely the number we find, corroborating the FCI nature of the state.

\textit{Structure factor.}
In the band projected models like tTMDs, there are two ways to define the structure factor $S(\bm{q})$. In this paper, convention in Ref.~\cite{zaklama2025structurefactortopologicalbound} is used, with the first-quantized expression $S(\bm{q})\equiv  \frac{1}{N_e} \langle \sum_{i,j} e^{-i\bm{q}\cdot(\bm{r}_i - \bm{r}_j)}\rangle$ that can be rewritten as
\begin{equation}
\begin{split}
        S(\bm{q}) 
        %&= \frac{1}{N_e} \langle \sum_{i,j} e^{-i\bm{q}\cdot(\bm{r}_i - \bm{r}_j)}\rangle\\
        &= \frac{1}{N_e}\left(\left\langle \sum_{i=j}e^{-i\bm{q}\cdot(\bm{r}_i - \bm{r}_i)}\right\rangle +\left\langle \sum_{i\neq j}e^{-i\bm{q}\cdot(\bm{r}_i - \bm{r}_j)}\right\rangle\right)\\
        &=1 + \frac{1}{N_e}S_2(\bm{q}).
\end{split}
\end{equation}
Here we averaged the structure factor with electron numbers $N_e$. The second part of the structure factor $S_2(\bm{q})$ is just normal ordered projected density-density correlation function if we assume the ground state is within the projected subspace $P\vert\Psi_{GS}\rangle = \vert\Psi_{GS}\rangle$.
\begin{equation}
\begin{split}
        S_2(\bm{q}) &= \left\langle \sum_{i\neq j}Pe^{-i\bm{r}_i \cdot \bm{q}} P e^{i\bm{r}_j \cdot \bm{q}}P\right\rangle = \langle:\tilde{\rho}(\bm{q}) \tilde{\rho}(-\bm{q}):\rangle\\
        &= \sum_{\bm{k},\bm{k}'}\sum_{pqrs}M_{pq}(\bm{k},\bm{q}) M_{rs}(\bm{k}',-\bm{q})\\
        &\langle c^\dagger_p(\bm{k}+\bm{q})c^\dagger_r(\bm{k}'-\bm{q})c_s(\bm{k}')c_q(\bm{k})\rangle, 
\end{split}
\end{equation}
where $p,q,r,s$ are band indices, and in our case span the lowest three bands.
The fully spin-polarized sector is assumed. And the interacting energy has the relation to $S_2(\bm{q})$,
\begin{equation}
    E_{\text{I}} = \sum_{\bm{q}} \frac{1}{2N\Omega_c} V(\bm{q}) S_2(\bm{q}).
\end{equation}
When $\bm{q} = 0$, we also subtract the background, so the final expression is
\begin{equation}
    S(\bm{q}) = 1 + \frac{1}{N_e}S_2(\bm{q}) - \delta_{\bm{q},0}N_e.
\end{equation}
For DSRG(2) calculation, only the value of $S_2(\bm{q})$ is renormalized. The renormalized operator is obtained first, and then we evaluate the expectation value using the relaxed wave function $\Phi'$. The details of derivation are given in the Supplemental Material.

\textit{Computing the quantum metric with interactions.}
We adopt a gauge-invariant formulation of the geometric tensor 
$\overline{B}_{ij}(\bm{k}) = \mathrm{Tr}\left[\overline{P}(\bm{k})\; \partial_i \overline{P}(\bm{k})\, \partial_j \overline{P}(\bm{k})\right]$ defined in terms of the projection matrix $\overline{P}(\bm{k})$ on the SRG lowest band. This quantity enters the density operator in the lowest band
% Here a new set of correlated orbitals are defined using a operator $\widehat{N}_1$ that is connected to noninteracting projection operator $P_{nm}(\bm{k})$,
%
\begin{equation}
\begin{split}
  \widehat{\overline{N}{}}_1 &= \sum_{\bm{k}}  e^{-A}c^\dagger_1(\bm{k})c_1(\bm{k})e^{A}\\
  &= \sum_{\bm{k}} \sum_{\alpha,\beta} e^{-A}P_{\alpha\beta}(\bm{k})\psi^\dagger_\alpha(\bm{k})\psi_\beta(\bm{k})e^A,  
\end{split}
\end{equation} 
where $\psi_\alpha(\bm{k})$ denote the Bloch orbitals of the non-interacting system.
%The matrix $P(\bm{k})$ has properties of projection operators  $P^2(\bm{k}) = P(\bm{k})$, and $\psi_\alpha(\bm{k})$ is the Bloch orbital operator defined in the whole system. 
For noninteracting cases $A=0$, diagonalizing $P(\bm{k})$ would give the active band wave functions when eigenvalues are one. 
%The matrix $P(\bm{k})$ can be used in calculating topological properties of the system as shown in the SM. 
Turning on the interactions, the $\widehat{N}_1$ operator gets renormalized by the DSRG: 
%denoted as $\widehat{\overline{N}}_1$: 
\begin{equation}\label{eq:density operator}
\begin{split}
        \widehat{\overline{N}}_1 &= \overline{N}_1 +\sum_{\bm{k},p,q} \overline{N_1}^{q}_{p}(\bm{k}) \{c^\dagger_p(\bm{k})c_q(\bm{k})\} + \ldots\\
        &=\overline{N}_1 + \sum_{\bm{k},\alpha,\beta} \overline{P}_{\alpha \beta}(\bm{k}) \{\psi^\dagger_\alpha(\bm{k}) \psi_\beta(\bm{k})\}+\ldots,
\end{split}
\end{equation}
where $\overline{N}_1$ is the expectation value of total electron charge of the lowest band of the new many-body SRG wavefunction $e^{A}\vert \Phi \rangle$. 
The matrix $\overline{N_1}^{q}_{p}(\bm{k})$ is related to our sought  $\overline{P}_{\alpha\beta}(\bm{k})$ by a rotation into the Bloch orbital basis. We define the eigenvector of the biggest eigenvalue of $\overline{P}_{\alpha\beta}(\bm{k})$ matrix as the correlated orbital. We note that this is not the unique way to define these orbitals. For example, natural orbitals which diagonalize the one-body reduced density matrix are commonly used to study strongly correlation in molecules. However, the band topology is most naturally defined by band projection operator, as it is thus reasonable to use correlated orbitals inherited from the noninteracting projection matrix.

\renewcommand{\emph}{\textit}
\bibliography{ref.bib}% Produces the bibliography via BibTeX.

@article{Brouder2007,
  title = {Exponential Localization of Wannier Functions in Insulators},
  author = {Brouder, Christian and Panati, Gianluca and Calandra, Matteo and Mourougane, Christophe and Marzari, Nicola},
  journal = {Phys. Rev. Lett.},
  volume = {98},
  issue = {4},
  pages = {046402},
  numpages = {4},
  year = {2007},
  month = {Jan},
  publisher = {American Physical Society},
  doi = {10.1103/PhysRevLett.98.046402},
  url = {https://link.aps.org/doi/10.1103/PhysRevLett.98.046402}
}

@article{tracecondition,
  title = {Band geometry of fractional topological insulators},
  author = {Roy, Rahul},
  journal = {Phys. Rev. B},
  volume = {90},
  issue = {16},
  pages = {165139},
  numpages = {7},
  year = {2014},
  month = {Oct},
  publisher = {American Physical Society},
  doi = {10.1103/PhysRevB.90.165139},
  url = {https://link.aps.org/doi/10.1103/PhysRevB.90.165139}
}

@article{HerzogArbeitman2022,
  title = {Superfluid Weight Bounds from Symmetry and Quantum Geometry in Flat Bands},
  author = {Herzog-Arbeitman, Jonah and Peri, Valerio and Schindler, Frank and Huber, Sebastian D. and Bernevig, B. Andrei},
  journal = {Phys. Rev. Lett.},
  volume = {128},
  issue = {8},
  pages = {087002},
  numpages = {8},
  year = {2022},
  month = {Feb},
  publisher = {American Physical Society},
  doi = {10.1103/PhysRevLett.128.087002},
  url = {https://link.aps.org/doi/10.1103/PhysRevLett.128.087002}
}

@article{Wilson1993,
  title = {Renormalization of Hamiltonians},
  author = {G\l{}azek, Stanis\l{}aw D. and Wilson, Kenneth G.},
  journal = {Phys. Rev. D},
  volume = {48},
  issue = {12},
  pages = {5863--5872},
  numpages = {0},
  year = {1993},
  month = {Dec},
  publisher = {American Physical Society},
  doi = {10.1103/PhysRevD.48.5863},
  url = {https://link.aps.org/doi/10.1103/PhysRevD.48.5863}
}

@article{Wegner1994,
author = {Wegner, Franz},
title = {Flow-equations for Hamiltonians},
journal = {Annalen der Physik},
volume = {506},
number = {2},
pages = {77-91},
doi = {https://doi.org/10.1002/andp.19945060203},
year = {1994}
}

@article{JiabinMulti,
  title = {Fractional Chern insulators versus nonmagnetic states in twisted bilayer ${\mathrm{MoTe}}_{2}$},
  author = {Yu, Jiabin and Herzog-Arbeitman, Jonah and Wang, Minxuan and Vafek, Oskar and Bernevig, B. Andrei and Regnault, Nicolas},
  journal = {Phys. Rev. B},
  volume = {109},
  issue = {4},
  pages = {045147},
  numpages = {37},
  year = {2024},
  month = {Jan},
  publisher = {American Physical Society},
  doi = {10.1103/PhysRevB.109.045147},
  url = {https://link.aps.org/doi/10.1103/PhysRevB.109.045147}
}

@article{XuPNAS,
author = {Cheng Xu  and Jiangxu Li  and Yong Xu  and Zhen Bi  and Yang Zhang },
title = {Maximally localized Wannier functions, interaction models, and fractional quantum anomalous Hall effect in twisted bilayer ${\mathrm{MoTe}}_{2}$},
journal = {Proceedings of the National Academy of Sciences},
volume = {121},
number = {8},
pages = {e2316749121},
year = {2024},
doi = {10.1073/pnas.2316749121},
URL = {https://www.pnas.org/doi/abs/10.1073/pnas.2316749121},
eprint = {https://www.pnas.org/doi/pdf/10.1073/pnas.2316749121},
}

@article{mixingLiangFu,
  title = {Band mixing in the quantum anomalous Hall regime of twisted semiconductor bilayers},
  author = {Abouelkomsan, Ahmed and Reddy, Aidan P. and Fu, Liang and Bergholtz, Emil J.},
  journal = {Phys. Rev. B},
  volume = {109},
  issue = {12},
  pages = {L121107},
  numpages = {6},
  year = {2024},
  month = {Mar},
  publisher = {American Physical Society},
  doi = {10.1103/PhysRevB.109.L121107},
  url = {https://link.aps.org/doi/10.1103/PhysRevB.109.L121107}
}

@article{wickd,
    author = {Evangelista, Francesco A.},
    title = {Automatic derivation of many-body theories based on general Fermi vacua},
    journal = {The Journal of Chemical Physics},
    volume = {157},
    number = {6},
    pages = {064111},
    year = {2022},
    month = {08},
    issn = {0021-9606},
    doi = {10.1063/5.0097858},
    url = {https://doi.org/10.1063/5.0097858},
}

@misc{zaklama2025structurefactortopologicalbound,
      title={Structure factor and topological bound of twisted bilayer semiconductors at fractional fillings}, 
      author={Timothy Zaklama and Di Luo and Liang Fu},
      year={2025},
      eprint={2411.03496},
      archivePrefix={arXiv},
      primaryClass={cond-mat.str-el},
      url={https://arxiv.org/abs/2411.03496}, 
}

@article{TopologicalBound2024,
  title = {Topological Bound on the Structure Factor},
  author = {Onishi, Yugo and Fu, Liang},
  journal = {Phys. Rev. Lett.},
  volume = {133},
  issue = {20},
  pages = {206602},
  numpages = {6},
  year = {2024},
  month = {Nov},
  publisher = {American Physical Society},
  doi = {10.1103/PhysRevLett.133.206602},
  url = {https://link.aps.org/doi/10.1103/PhysRevLett.133.206602}
}

@article{DSRG2014,
    author = {Evangelista, Francesco A.},
    title = {A driven similarity renormalization group approach to quantum many-body problems},
    journal = {The Journal of Chemical Physics},
    volume = {141},
    number = {5},
    pages = {054109},
    year = {2014},
    month = {08},
    issn = {0021-9606},
    doi = {10.1063/1.4890660},
    url = {https://doi.org/10.1063/1.4890660},
}

@article{annurev:DSRG,
   author = "Li, Chenyang and Evangelista, Francesco A.",
   title = "Multireference Theories of Electron Correlation Based on the Driven Similarity Renormalization Group", 
   journal= "Annual Review of Physical Chemistry",
   year = "2019",
   volume = "70",
   number = "Volume 70, 2019",
   pages = "245-273",
   doi = "https://doi.org/10.1146/annurev-physchem-042018-052416",
   url = "https://www.annualreviews.org/content/journals/10.1146/annurev-physchem-042018-052416",
   publisher = "Annual Reviews",
   issn = "1545-1593",
   type = "Journal Article",
  }

@article{DSRGexcited,
    author = {Li, Chenyang and Evangelista, Francesco A.},
    title = {Driven similarity renormalization group for excited states: A state-averaged perturbation theory},
    journal = {The Journal of Chemical Physics},
    volume = {148},
    number = {12},
    pages = {124106},
    year = {2018},
    month = {03},
    issn = {0021-9606},
    doi = {10.1063/1.5019793},
    url = {https://doi.org/10.1063/1.5019793},
}

@article{DSRG2,
    author = {Li, Chenyang and Evangelista, Francesco A.},
    title = {Towards numerically robust multireference theories: The driven similarity renormalization group truncated to one- and two-body operators},
    journal = {The Journal of Chemical Physics},
    volume = {144},
    number = {16},
    pages = {164114},
    year = {2016},
    month = {04},
    issn = {0021-9606},
    doi = {10.1063/1.4947218},
    url = {https://doi.org/10.1063/1.4947218},
}

@article{DSRGex2,
author = {Wang, Meng and Fang, Wei-Hai and Li, Chenyang},
title = {Assessment of State-Averaged Driven Similarity Renormalization Group on Vertical Excitation Energies: Optimal Flow Parameters and Applications to Nucleobases},
journal = {Journal of Chemical Theory and Computation},
volume = {19},
number = {1},
pages = {122-136},
year = {2023},
doi = {10.1021/acs.jctc.2c00966},
note ={PMID: 36534617},
URL = {https://doi.org/10.1021/acs.jctc.2c00966}
}

@Inbook{GNOMukherjee1995,
author="Mukherjee, D.",
editor="Schachinger, E.
and Mitter, H.
and Sormann, H.",
title="A Coupled Cluster Approach to the Electron Correlation Problem Using a Correlated Reference State",
bookTitle="Recent Progress in Many-Body Theories: Volume 4",
year="1995",
publisher="Springer US",
address="Boston, MA",
pages="127--133",
isbn="978-1-4615-1937-9",
doi="10.1007/978-1-4615-1937-9_12",
url="https://doi.org/10.1007/978-1-4615-1937-9_12"
}

@article{GNOMUKHERJEE1997561,
title = {Normal ordering and a Wick-like reduction theorem for fermions with respect to a multi-determinantal reference state},
journal = {Chemical Physics Letters},
volume = {274},
number = {5},
pages = {561-566},
year = {1997},
issn = {0009-2614},
doi = {https://doi.org/10.1016/S0009-2614(97)00714-8},
url = {https://www.sciencedirect.com/science/article/pii/S0009261497007148},
author = {Debashis Mukherjee}
}

@article{GNO1997,
    author = {Kutzelnigg, Werner and Mukherjee, Debashis},
    title = {Normal order and extended Wick theorem for a multiconfiguration reference wave function},
    journal = {The Journal of Chemical Physics},
    volume = {107},
    number = {2},
    pages = {432-449},
    year = {1997},
    month = {07},
    issn = {0021-9606},
    doi = {10.1063/1.474405},
    url = {https://doi.org/10.1063/1.474405},
}

@article{GNO2013,
title = {Generalized antisymmetric ordered products, generalized normal ordered products, ordered and ordinary cumulants and their use in many electron correlation problem},
journal = {Computational and Theoretical Chemistry},
volume = {1003},
pages = {62-70},
year = {2013},
note = {Reduced Density Matrices: A Simpler Approach to Many-Electron Problems?},
issn = {2210-271X},
doi = {https://doi.org/10.1016/j.comptc.2012.09.035},
url = {https://www.sciencedirect.com/science/article/pii/S2210271X12005439},
author = {Debalina Sinha and Rahul Maitra and Debashis Mukherjee}
}

@Article{FCICai2023,
author={Cai, Jiaqi
and Anderson, Eric
and Wang, Chong
and Zhang, Xiaowei
and Liu, Xiaoyu
and Holtzmann, William
and Zhang, Yinong
and Fan, Fengren
and Taniguchi, Takashi
and Watanabe, Kenji
and Ran, Ying
and Cao, Ting
and Fu, Liang
and Xiao, Di
and Yao, Wang
and Xu, Xiaodong},
title={Signatures of fractional quantum anomalous Hall states in twisted MoTe2},
journal={Nature},
year={2023},
month={Oct},
day={01},
volume={622},
number={7981},
pages={63-68},
issn={1476-4687},
doi={10.1038/s41586-023-06289-w},
url={https://doi.org/10.1038/s41586-023-06289-w}
}

@Article{FCIPark2023,
author={Park, Heonjoon
and Cai, Jiaqi
and Anderson, Eric
and Zhang, Yinong
and Zhu, Jiayi
and Liu, Xiaoyu
and Wang, Chong
and Holtzmann, William
and Hu, Chaowei
and Liu, Zhaoyu
and Taniguchi, Takashi
and Watanabe, Kenji
and Chu, Jiun-Haw
and Cao, Ting
and Fu, Liang
and Yao, Wang
and Chang, Cui-Zu
and Cobden, David
and Xiao, Di
and Xu, Xiaodong},
title={Observation of fractionally quantized anomalous Hall effect},
journal={Nature},
year={2023},
month={Oct},
day={01},
volume={622},
number={7981},
pages={74-79},
issn={1476-4687},
doi={10.1038/s41586-023-06536-0},
url={https://doi.org/10.1038/s41586-023-06536-0}
}

@Article{FCIZeng2023,
author={Zeng, Yihang
and Xia, Zhengchao
and Kang, Kaifei
and Zhu, Jiacheng
and Kn{\"u}ppel, Patrick
and Vaswani, Chirag
and Watanabe, Kenji
and Taniguchi, Takashi
and Mak, Kin Fai
and Shan, Jie},
title={Thermodynamic evidence of fractional Chern insulator in moir{\'e} MoTe2},
journal={Nature},
year={2023},
month={Oct},
day={01},
volume={622},
number={7981},
pages={69-73},
issn={1476-4687},
doi={10.1038/s41586-023-06452-3},
url={https://doi.org/10.1038/s41586-023-06452-3}
}

@Article{FCIJi2024,
author={Ji, Zhurun
and Park, Heonjoon
and Barber, Mark E.
and Hu, Chaowei
and Watanabe, Kenji
and Taniguchi, Takashi
and Chu, Jiun-Haw
and Xu, Xiaodong
and Shen, Zhi-Xun},
title={Local probe of bulk and edge states in a fractional Chern insulator},
journal={Nature},
year={2024},
month={Nov},
day={01},
volume={635},
number={8039},
pages={578-583},
issn={1476-4687},
doi={10.1038/s41586-024-08092-7},
url={https://doi.org/10.1038/s41586-024-08092-7}
}

@Article{FCIWang2025,
author={Wang, Yiping
and Choe, Jeongheon
and Anderson, Eric
and Li, Weijie
and Ingham, Julian
and Arsenault, Eric A.
and Li, Yiliu
and Hu, Xiaodong
and Taniguchi, Takashi
and Watanabe, Kenji
and Roy, Xavier
and Basov, Dmitri
and Xiao, Di
and Queiroz, Raquel
and Hone, James C.
and Xu, Xiaodong
and Zhu, X.-Y.},
title={Hidden states and dynamics of fractional fillings in twisted MoTe2 bilayers},
journal={Nature},
year={2025},
month={May},
day={01},
volume={641},
number={8065},
pages={1149-1155},
doi={10.1038/s41586-025-08954-8},
url={https://doi.org/10.1038/s41586-025-08954-8}
}

@Article{SCXia2025,
author={Xia, Yiyu
and Han, Zhongdong
and Watanabe, Kenji
and Taniguchi, Takashi
and Shan, Jie
and Mak, Kin Fai},
title={Superconductivity in twisted bilayer WSe2},
journal={Nature},
year={2025},
month={Jan},
day={01},
volume={637},
number={8047},
pages={833-838},
issn={1476-4687},
doi={10.1038/s41586-024-08116-2},
url={https://doi.org/10.1038/s41586-024-08116-2}
}

@Article{SCGuo2025,
author={Guo, Yinjie
and Pack, Jordan
and Swann, Joshua
and Holtzman, Luke
and Cothrine, Matthew
and Watanabe, Kenji
and Taniguchi, Takashi
and Mandrus, David G.
and Barmak, Katayun
and Hone, James
and Millis, Andrew J.
and Pasupathy, Abhay
and Dean, Cory R.},
title={Superconductivity in 5.0{\textdegree} twisted bilayer WSe2},
journal={Nature},
year={2025},
month={Jan},
day={01},
volume={637},
number={8047},
pages={839-845},
issn={1476-4687},
doi={10.1038/s41586-024-08381-1},
url={https://doi.org/10.1038/s41586-024-08381-1}
}

@article{LiangFuED,
  title = {Anomalous Hall metal and fractional Chern insulator in twisted transition metal dichalcogenides},
  author = {Cr\'epel, Valentin and Fu, Liang},
  journal = {Phys. Rev. B},
  volume = {107},
  issue = {20},
  pages = {L201109},
  numpages = {5},
  year = {2023},
  month = {May},
  publisher = {American Physical Society},
  doi = {10.1103/PhysRevB.107.L201109},
  url = {https://link.aps.org/doi/10.1103/PhysRevB.107.L201109}
}

@article{VPED,
  title = {Spontaneous fractional Chern insulators in transition metal dichalcogenide moir\'e superlattices},
  author = {Li, Heqiu and Kumar, Umesh and Sun, Kai and Lin, Shi-Zeng},
  journal = {Phys. Rev. Res.},
  volume = {3},
  issue = {3},
  pages = {L032070},
  numpages = {6},
  year = {2021},
  month = {Sep},
  publisher = {American Physical Society},
  doi = {10.1103/PhysRevResearch.3.L032070},
  url = {https://link.aps.org/doi/10.1103/PhysRevResearch.3.L032070}
}

@article{PressureED,
  title = {Pressure-enhanced fractional Chern insulators along a magic line in moir\'e transition metal dichalcogenides},
  author = {Morales-Dur\'an, Nicol\'as and Wang, Jie and Schleder, Gabriel R. and Angeli, Mattia and Zhu, Ziyan and Kaxiras, Efthimios and Repellin, C\'ecile and Cano, Jennifer},
  journal = {Phys. Rev. Res.},
  volume = {5},
  issue = {3},
  pages = {L032022},
  numpages = {6},
  year = {2023},
  month = {Aug},
  publisher = {American Physical Society},
  doi = {10.1103/PhysRevResearch.5.L032022},
  url = {https://link.aps.org/doi/10.1103/PhysRevResearch.5.L032022}
}

@article{displacementED,
  title = {Topological quantum phase transitions driven by a displacement field in twisted ${\mathrm{MoTe}}_{2}$ bilayers},
  author = {Sharma, Prakash and Peng, Yang and Sheng, D. N.},
  journal = {Phys. Rev. B},
  volume = {110},
  issue = {12},
  pages = {125142},
  numpages = {20},
  year = {2024},
  month = {Sep},
  publisher = {American Physical Society},
  doi = {10.1103/PhysRevB.110.125142},
  url = {https://link.aps.org/doi/10.1103/PhysRevB.110.125142}
}

@article{VPtheoryQiu,
  title = {Interaction-Driven Topological Phase Diagram of Twisted Bilayer ${\mathrm{MoTe}}_{2}$},
  author = {Qiu, Wen-Xuan and Li, Bohao and Luo, Xun-Jiang and Wu, Fengcheng},
  journal = {Phys. Rev. X},
  volume = {13},
  issue = {4},
  pages = {041026},
  numpages = {17},
  year = {2023},
  month = {Nov},
  publisher = {American Physical Society},
  doi = {10.1103/PhysRevX.13.041026},
  url = {https://link.aps.org/doi/10.1103/PhysRevX.13.041026}
}

@article{TMDmodel,
  title = {Fractional Chern Insulator in Twisted Bilayer ${\mathrm{MoTe}}_{2}$},
  author = {Wang, Chong and Zhang, Xiao-Wei and Liu, Xiaoyu and He, Yuchi and Xu, Xiaodong and Ran, Ying and Cao, Ting and Xiao, Di},
  journal = {Phys. Rev. Lett.},
  volume = {132},
  issue = {3},
  pages = {036501},
  numpages = {6},
  year = {2024},
  month = {Jan},
  publisher = {American Physical Society},
  doi = {10.1103/PhysRevLett.132.036501},
  url = {https://link.aps.org/doi/10.1103/PhysRevLett.132.036501}
}

@article{TopoAnderson,
  title = {Superconductivity in Twisted ${\mathrm{WSe}}_{2}$ from Topology-Induced Quantum Fluctuations},
  author = {Xie, Fang and Chen, Lei and Sur, Shouvik and Fang, Yuan and Cano, Jennifer and Si, Qimiao},
  journal = {Phys. Rev. Lett.},
  volume = {134},
  issue = {13},
  pages = {136503},
  numpages = {6},
  year = {2025},
  month = {Mar},
  publisher = {American Physical Society},
  doi = {10.1103/PhysRevLett.134.136503},
  url = {https://link.aps.org/doi/10.1103/PhysRevLett.134.136503}
}

@misc{xie2025kondolatticephenomenologytwistedbilayer,
      title={Kondo-lattice phenomenology of twisted bilayer WSe$_2$ from compact molecular orbitals of topological bands}, 
      author={Fang Xie and Chenyuan Li and Jennifer Cano and Qimiao Si},
      year={2025},
      eprint={2503.21769},
      archivePrefix={arXiv},
      primaryClass={cond-mat.str-el},
      url={https://arxiv.org/abs/2503.21769}, 
}

@misc{li2025deeplearningshedslight,
      title={Deep Learning Sheds Light on Integer and Fractional Topological Insulators}, 
      author={Xiang Li and Yixiao Chen and Bohao Li and Haoxiang Chen and Fengcheng Wu and Ji Chen and Weiluo Ren},
      year={2025},
      eprint={2503.11756},
      archivePrefix={arXiv},
      primaryClass={cond-mat.str-el},
      url={https://arxiv.org/abs/2503.11756}, 
}

@article{Regnault2011,
  title = {Fractional Chern Insulator},
  author = {Regnault, Nicolas and Bernevig, B. Andrei},
  journal = {Phys. Rev. X},
  volume = {1},
  OPTissue = {1},
  pages = {021014},
  year = {2011},
  doi = {10.1103/PhysRevX.1.021014},
  url = {https://link.aps.org/doi/10.1103/PhysRevX.1.021014}
}

@Article{Resta2011,
author={Resta, R.},
title={The insulating state of matter: a geometrical theory},
journal={The European Physical Journal B},
year={2011},
month={Jan},
day={01},
volume={79},
number={2},
pages={121-137},
doi={10.1140/epjb/e2010-10874-4},
url={https://doi.org/10.1140/epjb/e2010-10874-4}
}

@Article{Yu2025,
author={Yu, Jiabin
and Bernevig, B. Andrei
and Queiroz, Raquel
and Rossi, Enrico
and T{\"o}rm{\"a}, P{\"a}ivi
and Yang, Bohm-Jung},
title={Quantum geometry in quantum materials},
journal={npj Quantum Materials},
year={2025},
month={Oct},
day={10},
volume={10},
number={1},
pages={101},
doi={10.1038/s41535-025-00801-3},
url={https://doi.org/10.1038/s41535-025-00801-3}
}

@article{ctbg2020,
  title = {Fractional Chern insulator states in twisted bilayer graphene: An analytical approach},
  author = {Ledwith, Patrick J. and Tarnopolsky, Grigory and Khalaf, Eslam and Vishwanath, Ashvin},
  journal = {Phys. Rev. Res.},
  volume = {2},
  issue = {2},
  pages = {023237},
  numpages = {12},
  year = {2020},
  month = {May},
  publisher = {American Physical Society},
  doi = {10.1103/PhysRevResearch.2.023237},
  url = {https://link.aps.org/doi/10.1103/PhysRevResearch.2.023237}
}

@article{ctbg2021,
  title = {Exact Landau Level Description of Geometry and Interaction in a Flatband},
  author = {Wang, Jie and Cano, Jennifer and Millis, Andrew J. and Liu, Zhao and Yang, Bo},
  journal = {Phys. Rev. Lett.},
  volume = {127},
  issue = {24},
  pages = {246403},
  numpages = {6},
  year = {2021},
  month = {Dec},
  publisher = {American Physical Society},
  doi = {10.1103/PhysRevLett.127.246403},
  url = {https://link.aps.org/doi/10.1103/PhysRevLett.127.246403}
}

@article{ctbg2022,
  title = {Family of Ideal Chern Flatbands with Arbitrary Chern Number in Chiral Twisted Graphene Multilayers},
  author = {Ledwith, Patrick J. and Vishwanath, Ashvin and Khalaf, Eslam},
  journal = {Phys. Rev. Lett.},
  volume = {128},
  issue = {17},
  pages = {176404},
  numpages = {7},
  year = {2022},
  month = {Apr},
  publisher = {American Physical Society},
  doi = {10.1103/PhysRevLett.128.176404},
  url = {https://link.aps.org/doi/10.1103/PhysRevLett.128.176404}
}

\clearpage
\setcounter{equation}{0}
\setcounter{figure}{0}
\setcounter{table}{0}

\makeatletter
\renewcommand{\theequation}{S\arabic{equation}}
\renewcommand{\thefigure}{S\arabic{figure}}
\renewcommand{\thetable}{S\arabic{table}}

% Make hyperref anchors unique as well:
\renewcommand{\theHequation}{S\arabic{equation}}
\renewcommand{\theHfigure}{S\arabic{figure}}
\renewcommand{\theHtable}{S\arabic{table}}

\renewcommand{\bibnumfmt}[1]{[#1]}
\renewcommand{\citenumfont}[1]{#1}

\setcounter{section}{0}
\renewcommand\thesection{\Roman{section}}

\onecolumngrid

%\begin{widetext}
\begin{center}
	\Large{\textbf{Supplementary Materials for ``\thistitle"}}
\end{center}
%\end{widetext}

\section{Details of DSRG(2) calculations}
To begin with, we introduce several conventions that will be used. The many-body operators will be expressed in the form as 
\begin{equation}\label{eq:operator}
     \widehat{O} =O_0 + (O_1)^s_r \{\widehat{a}^{r}_s\}+\frac{1}{(2!)^2}(O_{2})^{rs}_{pq}\{\widehat{a}^{pq}_{rs}\} + \ldots,
\end{equation}
where the prefactor of normal ordered n-body operator should be $\frac{1}{(n!)^2}$, as the n-body tensors $(O_n)^{pqrs\ldots}_{ijkl\ldots}$ are anti-symmetrized separately in the upper and lower indices. Here, Einstein summation is assumed. The braces stand for generalized normal ordering (GNO), and the operators are represented as abbreviation of $\widehat{a}^{pq\ldots}_{rs\ldots} \equiv a^\dagger_{p} a^\dagger_q \ldots a_{s} a_{r}$.

\subsection{Generalized normal ordering}
The normal ordering, by definition, requires $\langle \Phi\vert \{\widehat{O}\}\vert\Phi\rangle = 0$, for an arbitrary many-body state $\Phi$, where $\{\widehat{O}\}$ represents generalized normal ordering with respect to the reference state. This condition should be applied when expanding many-body operators. The package \textsc{Wick\&d} \cite{wickd} will apply generalized Wick theorem and GNO automatically. The density matrices of $\Phi$, $\gamma^{pqr\ldots}_{xyz\ldots} = \langle\Phi\vert a^\dagger_p a^\dagger_q a^\dagger_r \ldots a_z a_y a_x \vert\Phi\rangle$, need to be input to the program. For example, the one-particle $\gamma_1$ and one-hole density matrices $\eta_1$ are
\begin{equation}
    \begin{split}
        \gamma^p_q &= \langle \Phi\vert a^\dagger_p a_q\vert\Phi\rangle\\
        \eta^{p}_q &= \langle \Phi\vert a_q a^\dagger_p\vert\Phi\rangle = \delta^{p}_q - \gamma^p_q.
    \end{split}
\end{equation}
For multi-leg contractions, higher order density matrices like two-body cumulants $\lambda_2$ are needed
\begin{equation}
    \lambda^{pq}_{rs} = \gamma^{pq}_{rs} - \gamma^{p}_r\gamma^{q}_s + \gamma^p_{s} \gamma^q_r.
\end{equation}

For example, the one-body operators can be normal ordered as 
\begin{equation}
    \{\widehat{a}^p_q\} = \widehat{a}^p_q - \gamma^p_q,
\end{equation}
and GNO of the two-body operators can be expressed as
\begin{equation}
    \{\widehat{a}^{xy}_{uv}\} = \widehat{a}^{xy}_{uv}- \{\widehat{a}^{x}_u\} \gamma^y_v  - \{\widehat{a}^{y}_v\} \gamma^x_u+ \{\widehat{a}^{x}_v\} \gamma^y_u+ \{\widehat{a}^{y}_u\} \gamma^x_v - (\gamma^x_u \gamma^y_v - \gamma^x_v \gamma^y_u + \lambda^{xy}_{uv}).
\end{equation}
For details, we recommend readers to check Ref.~\cite{wickd}.

For degenerate ground states, we only need to average the n-body density matrices among $N_{GS}$ ground states,
\begin{equation}
    \overline{\gamma} = \sum_{i\in \text{GS}} \frac{1}{N_{GS}}\gamma_i,
\end{equation}
and the cumulants can be obtained using averaged density matrices, leaving GNO expressions unchanged.

\subsection{Linear approximation of BCH expansions}
As mentioned in the main text, we approximate the BCH expansion as
\begin{equation}\label{sup:bch}
\begin{split}
        \overline{H} &= \sum_{k=0} B_k,\\
        B_k & = \frac{1}{k} [B_{k-1},A]_{0,1,2},
\end{split}
\end{equation}
where we have truncated the operators up to normal ordered two-body operators. This guarantees that we have the same expressions for the commutator for every iteration, and the expansion can be set to terminate when the operator absolute difference is smaller than a threshold $\tau$,
\begin{equation}\label{eq:dsrg_converge}
    |\sum^{p+1}_{k=0} B_k - \sum^p_{k=0} B_k| < \tau.
\end{equation}
This approximation is called LDSRG(2) \cite{DSRG2}. Because we do not stop the expansion in a finite step, this expansion is thought to be non-perturbative. 

\subsection{Driven similarity renormalization group}
We treat the lowest band in the hole Hamiltonian as the active band, forming an active Fock space. The lowest active band will be labeled by indices $\{i,j,k,l\}$, and all bands are labeled by indices $\{a,b,c,d,p,q,r,s\}$ including the lowest band. The goal is to eliminate non-diagonal many-body couplings $ H^{a}_i,~ H^{i}_{a},~ \frac{1}{4}H^{ab}_{ij} ,~\text{and} ~\frac{1}{4}H^{ij}_{ab}$, when not all indices of each tensor belong to the active band. So we introduce an anti-hermitian operator $A = T-T^\dagger$ to eliminate the non-diagonals $\overline{H} = e^{-A } H e^A$. The renormalized Hamiltonian is obtained from BCH expansion from Eq.~\ref{sup:bch}.
%, and Eq.~\ref{eq:recursion}. 
The expressions can be found in \cite{DSRG2}, or can be obtained from \textsc{Wick\&d} \cite{wickd}.

We briefly introduce the method to recursively find the tensors $T^{ij\ldots}_{ab\ldots}$.
\begin{equation}
\begin{split}
        &T = T^{i}_a\{\widehat{a}^a_i\} + \frac{1}{4}T^{ij}_{ab}\{\widehat{a}^{ab}_{ij}\},\\
        &T^{ij\ldots}_{ab\ldots} = 0,~\text{when all indices are in active space.}
\end{split}
\end{equation}
The standard way \cite{DSRG2,annurev:DSRG} to find tensors $T^{ij\ldots}_{ab\ldots}$ is to introduce a source term $R(s)$ to drive the non-diagonal Hamiltonian $H_{\text{N}}$,
\begin{equation}\label{eq:source}
\begin{split}&\left[\overline{H}^{ij\ldots}_{ab\ldots}\right]_{\text{N}}(s) = R^{ij\ldots}_{ab\ldots}(s)\\
 &R^{ij\ldots}_{ab\ldots}(s) = \left\{\left[\overline{H}^{ij\ldots}_{ab\ldots}\right]_{\text{N}}(s) + T^{ij\ldots}_{ab \ldots}(s) \Delta^{ij\ldots}_{ab \ldots}\right \}e^{-s\left(\Delta^{ij\ldots}_{ab \ldots}\right)^2},
\end{split}
\end{equation}
where $\Delta^{ij\ldots}_{ab \ldots} = H^{i}_{i}+H^{j}_{j}+\ldots-H^{a}_a-H^b_b-\ldots$. Importantly, the tensor $H^{p}_q$ is not the one-body part of our Hamiltonian but the Fock operator due to the normal ordering. The reason to choose the form of $R(s)$ in Eq.~\ref{eq:source} is to mimic single-Slater determinant SRG flow in order to avoid degenerate band-crossing state (usually called intruder state). However, in the case of $\text{tMoTe}_2$, the band crossing is not present. The parameter $s$ serves as a smooth energy scale. The source operator has the boundary condition that $R(0) =H_{\text{N}}$, and $R(\infty) = 0$. In order to avoid numerical instabilities, we tested that taking the parameter as large as $s = 0.1$ has saturated the DSRG energies. 

The recursive equation can be derived from Eq.~\ref{eq:source},
\begin{equation}
    [T^{ij\ldots}_{ab\ldots}]_{\text{new}} = \left(\overline{H}^{ij\ldots}_{ab\ldots}+[T^{ij\ldots}_{ab\ldots}]_{\text{old}}\Delta^{ij\ldots}_{ab\ldots}\right)\frac{1-e^{-s\left(\Delta^{ij\ldots}_{ab\ldots}
\right)^2}}{\Delta^{ij\ldots}_{ab\ldots}},
\end{equation}
which is similar to fixed-point equations $x = f(x)$. In practice, the linear mixing method is applied to accelerate the convergence.

\subsection{Reference relaxation}
After obtaining the renormalized Hamiltonian $\overline{H}$, we diagonalize it within the active space. This process updates the original reference ground state, $\Phi \rightarrow \Phi'$, and yields a new set of density matrices, $\gamma \rightarrow \gamma'$. Importantly, the original unitary transformation $e^A$, which eliminates the non-diagonal components of the Hamiltonian, remains unchanged. The resulting ground state is given by $e^A \vert\Phi'\rangle$. Hence, The renormalized operator $\widehat{\overline{O}}$ is evaluated using the unitary operator $e^{A}$, but the expectation value needs to be calculated with respect to new reference $\langle \Phi'\vert \widehat{\overline{O}}\vert\Phi'\rangle$.

Because $\overline{H}$ is expressed in GNO form as shown in Eq.~\ref{eq:operator}, we need to recover it to the original one-body and two-body form without normal ordering in order to be input into the ED program. The expression is obtained simply by following the rules of GNO,
\begin{equation}\label{eq:relaxation}
\begin{split}
        H_0&= \overline{E}_0 -\overline{H}^{i}_j \gamma^{j}_i + \frac{1}{2} \overline{H}^{kl}_{ij} \gamma^{i}_k \gamma^j_l - \frac{1}{4}\overline{H}^{kl}_{ij} \lambda^{ij}_{kl},\\
        H_1 &=(\overline{H}^{p}_{q} - \gamma^i_{j} \overline{H}^{pj}_{qi})\widehat{a}^{q}_p,\\
        H_2 &= \frac{1}{4}\overline{H}^{ps}_{qr}\widehat{a}^{qr}_{ps}.
\end{split}
\end{equation}

\section{Static properties}

\subsection{Normal ordered expressions}
Taking the total charge operator within the lowest band $\widehat{N}_1$ as an example, we begin with full electron charges residing in the lowest band,
\begin{equation}
    \widehat{N}_1 = \sum_{\bm{k}}\langle c^\dagger_1(\bm{k})c_1(\bm{k})\rangle + \sum_{\bm{k}}\{c^\dagger_1(\bm{k})c_1(\bm{k})\}.
\end{equation}

After using BCH and GNO to expand the operator, the renormalized operator $\widehat{\overline{N}}_1$ is then expressed in GNO form with respect to state $\Phi$,

\begin{equation}\label{charge_density}
\begin{split}
        \widehat{\overline{N}}_1 =& e^{-A} \widehat{N}_1 e^{A} = \overline{N}_1 +\sum_{\bm{k}} \overline{N_1}^{q}_{p}(\bm{k}) \{c^\dagger_p(\bm{k})c_q(\bm{k})\}\\&+ \sum_{\bm{k},\bm{k}',\bm{q}}\frac{1}{4}(\overline{N}_1)^{rs}_{pq}(\bm{k},\bm{k}',\bm{q})\{c^\dagger_p(\bm{k}+\bm{q}) c^\dagger_q(\bm{k}'-\bm{q}) c_s(\bm{k}') c_r(\bm{k}) \} + \ldots,
\end{split}
\end{equation}
where  $\overline{N}_1$ is the expectation value of total electron charge of lowest band of the new many-body wave function $e^{A}\vert \Phi \rangle$. Here, the momentum conservation is not explicitly implemented, although it schematically shown in Eq.~\ref{charge_density}. In practice, we wrap the band indices and the momentum indices together, and make sure all the tensors are properly anti-symmetrized. To evaluate the expectation value of $\widehat{N}_1$ of relaxed reference $\Phi'$, it is similar to Eq.~\ref{eq:relaxation} to get non-GNO operators, and use new density matrices $\gamma'$ to evaluate the expectation values.

\subsection{Evaluation of structure factor}
In the band projected models like tTMDs, there are two ways to define the structure factor $S(\bm{q})$. In this paper, convention in Ref.~\cite{zaklama2025structurefactortopologicalbound} is used. The structure factor in the first-quantized expression is
\begin{equation}
\begin{split}
        S(\bm{q}) &= \frac{1}{N_e} \langle \sum_{i,j} e^{-i\bm{q}\cdot(\bm{r}_i - \bm{r}_j)}\rangle\\
        &= \frac{1}{N_e}\left(\langle \sum_{i=j}e^{-i\bm{q}\cdot(\bm{r}_i - \bm{r}_j)}\rangle +\langle \sum_{i\neq j}e^{-i\bm{q}\cdot(\bm{r}_i - \bm{r}_j)}\rangle\right)\\
        &=1 + \frac{1}{N_e}S_2(\bm{q}).
\end{split}
\end{equation}
Here we averaged the structure factor with electron numbers $N_e$. The second part of the structure factor $S_2(\bm{q})$ is just normal ordered projected density-density correlation function if we assume the ground state is within the projected subspace $P\vert\Psi_{GS}\rangle = \vert\Psi_{GS}\rangle$.
\begin{equation}
\begin{split}
        S_2(\bm{q}) &= \langle \sum_{i\neq j}Pe^{-i\bm{r}_i \cdot \bm{q}} P e^{i\bm{r}_j \cdot \bm{q}}P\rangle = \langle:\overline{\rho}(\bm{q}) \overline{\rho}(-\bm{q}):\rangle\\
        &= \sum_{\bm{k},\bm{k}'}\sum_{pqrs}M_{pq}(\bm{k},\bm{q}) M_{rs}(\bm{k}',-\bm{q})\\
        &\langle c^\dagger_p(\bm{k}+\bm{q})c^\dagger_r(\bm{k}'-\bm{q})c_s(\bm{k}')c_q(\bm{k})\rangle, 
\end{split}
\end{equation}
where $p,q,r,s$ are band indices, and in our case span the lowest three bands.
The fully spin-polarized sector is assumed. And the interacting energy has the relation to $S_2(\bm{q})$,
\begin{equation}
    E_{\text{I}} = \sum_{\bm{q}} \frac{1}{2N\Omega_c} V(\bm{q}) S_2(\bm{q}).
\end{equation}
When $\bm{q} = 0$, we also subtract the background, so the final expression is
\begin{equation}
    S(\bm{q}) = 1 + \frac{1}{N_e}S_2(\bm{q}) - \delta_{\bm{q},0}N_e.
\end{equation}
For DSRG(2) calculation, only the value of $S_2(\bm{q})$ is renormalized. Similar to the procedure described in Eq.~\ref{charge_density}, the renormalized operator is obtained first, and then we evaluate the expectation value using the relaxed wave function $\Phi'$.

\section{Berry curvature and quantum geometry}
Quantum geometric tensor (QGT) $\mathcal{B}_{ij}$ can be calculated by a gauge independent method \cite{Brouder2007,HerzogArbeitman2022},
\begin{equation}
    \mathcal{B}_{ij}(\bm{k}) = \mathrm{tr}\left[P(\bm{k}) \partial_i P(\bm{k}) \partial_j P(\bm{k})\right],
\end{equation}
where $P(\bm{k})$ is the $m$th band projection matrix:
\begin{equation}
        [P(\bm{k})]_{\bm{Q}_l\bm{Q}'_l} = u_{\bm{Q}_lm}(\bm{k}) u^*_{\bm{Q}'_lm}(\bm{k}).
\end{equation}
The QGT can be decomposed into real and imaginary part $Q_{ij} = g_{ij} + \frac{i}{2}f_{ij}$, where $g_{ij} = g_{ji}$ and $f_{ij} = -f_{ji}$. The real symmetric part is the Fubini-Study metric, and the Berry curvature in current basis is defined as $F(\bm{k}) = f_{21}$.

Chern number $\mathrm{Ch}$ can be evaluated by integrating the Berry curvature over the first Brillouin zone,
\begin{equation}
        2\pi \mathrm{Ch} =  \int_{BZ} d^2\bm{k}~ F(\bm{k}).
\end{equation}

\begin{figure}
    \centering
    \includegraphics[width=0.5\linewidth]{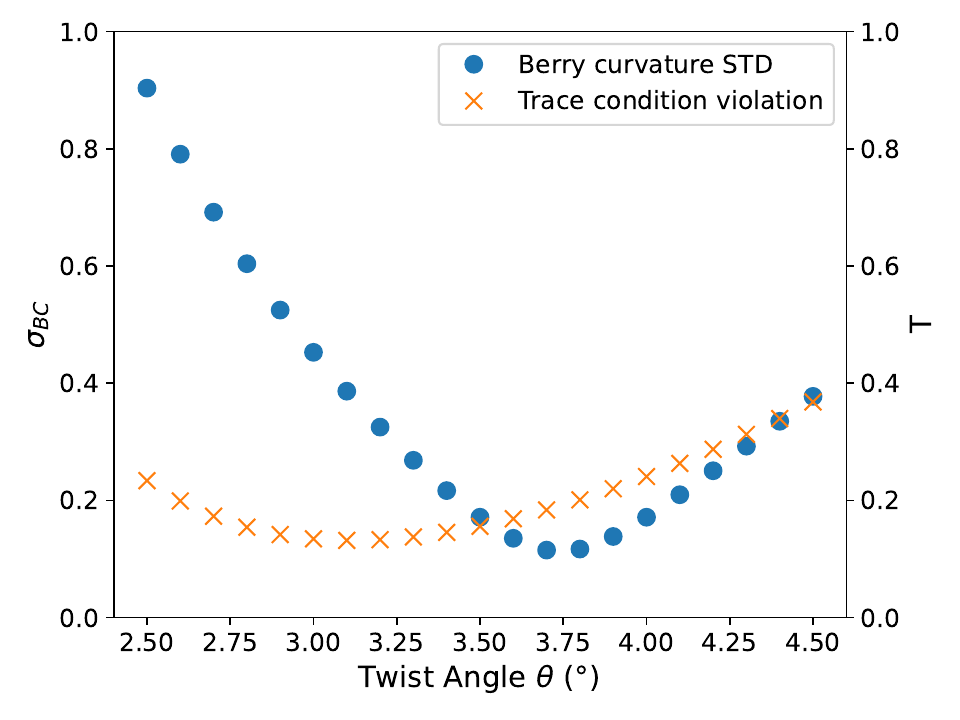}
    \caption{Berry curvature standard deviation $\sigma_{BC}$ and the violation of trace condition $T$ for the lowest bands of noninteracting tMoTe$_2$. The horizontal axis is twist angle $\theta$ ranging from $2.5\degree $ to $4.5\degree$. These quantities are evaluated in a $20\times20$ meshgrid. }
    \label{fig:std_and_T_nonint}
\end{figure}

To measure how close is the band to a Landau level, we evaluate the Berry curvature standard deviation $\sigma_{BC}$ and trace condition violation $T$. The Berry curvature standard deviation is defined as 
\begin{equation}
    \sigma_{BC} = \sqrt{\left [\frac{S^2_{BZ}}{4 \pi^2}  \left<F^2\right> - \mathrm{Ch}^{2} \right ]},
\end{equation}
where the bracket means average over the Brillouin zone.
We also define the trace condition violation as
\begin{equation}
    T = \int_{BZ} d^2\bm{k}\frac{ \mathrm{tr} ~ g(\bm{k})}{2\pi} - \left|\mathrm{Ch}\right|.
\end{equation}
There is a rigorous trace inequality \cite{tracecondition}, $\mathrm{tr}g(\bm{k}) = g_{xx}(\bm{k}) +g_{yy}(\bm{k})\geq |F(\bm{k})|$, and when the inequality saturates, the band is called ideal Chern band. $T$ is an overall quantification of violation of trace condition. In Fig.~\ref{fig:std_and_T_nonint}, we show $\sigma_{BC}$ and $T$ with respect to the twist angle $\theta$ for noninteracting tMoTe$_2$ in a $20\times20$ meshgrid, where $\sigma_{BC}$ is minimized around $3.7\degree$, while $\theta$ is smaller to take the lowest value of $T$ around $3.2\degree$. 

In the main text Fig.~\ref{fig:berry}, the trace condition violation is larger than that of Fig.~\ref{fig:std_and_T_nonint}, we checked and found this is purely size effects. In the interacting calculations, a small system size of $3\times9$ was used. The coarse size of the k-mesh means the discrete derivatives entering the expression for quantum metric are less accurate than when computing on a fine mesh in Fig.~\ref{fig:std_and_T_nonint}. $T$ for interacting systems is shown for comparison but not to compare with absolute value at larger systems.

\end{document}